\documentclass{iau307}
\usepackage{graphicx}
\usepackage{natbib}
\usepackage{url}
\usepackage{dtklogos}
\usepackage{caption}
\usepackage{subcaption}
\usepackage{amsmath,amssymb}
\bibpunct{(}{)}{;}{a}{}{,}

\newcommand{\zav}[1]{\left(#1\right)}

\newcommand\intvidpo{\!\!\int\limits_{\begin{array}{c}\text{\scriptsize
visible}\\[-2mm]\text{\scriptsize surface}\end{array}}\!\!}

\newlength\staretab

\makeatletter
\def\sgn{\mathop{\operator@font sgn}\nolimits}
\makeatother

\DeclareMathAlphabet{\mathsc}{OT1}{cmr}{m}{sc}
\def\testbx{bx}%
\DeclareRobustCommand{\ion}[2]{%
\relax\ifmmode
\ifx\testbx\f@series
{\mathbf{#1\,\mathsc{#2}}}\else
{\mathrm{#1\,\mathsc{#2}}}\fi
\else\textup{#1\,{\mdseries\textsc{#2}}}%
\fi}

\title[Unraveling the variability of $\sigma$ Ori E] 
{Unraveling the variability of $\sigma$ Ori E}

\author[Oksala et al.]   
{M.~E.~Oksala$^1$, O.~Kochukhov$^2$, J.~Krti\v{c}ka$^3$, M.~Prv\'{a}k$^3$, \\
 \and Z.Mikul\'{a}\v{s}ek$^3$}

\affiliation{$^1$Astronomical Institute, Academy of Sciences of the Czech Republic, Fricova 298, 251 65 Ond\v{r}ejov, Czech Republic \\ email: {\tt meo@udel.edu} \\[\affilskip]
$^2$Department of Physics and Astronomy, Uppsala University, Box 516, Uppsala 75120, Sweden\\
[\affilskip]
$^3$Institute of Theoretical Physics and Astrophysics, Masaryk University, 611 37 Brno, Czech Republic
}{

\pubyear{2014}
\volume{307} 
\pagerange{}
\setcounter{page}{1}
\jname{New windows on massive stars: asteroseismology, interferometry, and spectropolarimetry}
\editors{G. Meynet, C. Georgy, J.H. Groh \& Ph. Stee, eds.}

\begin{document}

\maketitle

\begin{abstract}
$\sigma$ Ori E (HD 37479) is the prototypical helium-strong star shown to harbor a strong magnetic field, as well as a magnetosphere consisting of two clouds of plasma.  The observed optical ($ubvy$) light curve of $\sigma$ Ori E is dominated by eclipse features due to circumstellar material, however, there remain additional features unexplained by the Rigidly Rotating Magnetosphere (RRM) model of Townsend \& Owocki.  Using the technique of magnetic Doppler imaging (MDI), spectropolarimetric observations of $\sigma$ Ori E are used to produce maps of both the magnetic field topology and various elemental abundance distributions.  We also present an analysis utilizing these computed MDI maps in conjunction with NLTE TLUSTY models to study the optical brightness variability of this star arising from surface inhomogeneities.  It has been suggested that this physical phenomena may be responsible for the light curve inconsistencies between the model and observations.
\end{abstract}

\firstsection 
\section{Introduction}

The B2Vp star $\sigma$ Ori E (HD\,37479) has long been known as the prototypical He-strong magnetic Bp star.  
Because of its status, it has been one of the most well studied of such stars, exhibiting numerous types of observed variability all modulated on the stellar rotation period.  
The combination of high rotation speed ($\sim40\%\,v_{\rm{crit}}$) and a strong, global magnetic field ($B_{\rm{p}} \sim 10$kG) makes $\sigma$ Ori E an excellent 
laboratory to study the interaction of magnetism, rotation, and mass-loss; three essential parameters to determine the stellar evolutionary course.   
With this enigmatic object in mind, \citet{Townsend:2005aa} developed their Rigidly Rotating Magnetosphere (RRM) model to describe 
the circumstellar environment of a oblique magnetic rotator in which rapid rotation keeps magnetically trapped material supported against gravity via centrifugal forces,
with material accumulating at the intersection between the rotational and magnetic equators.  

This analytical model, applied to the specific case of $\sigma$ Ori E by \citet{Townsend:2005ab}, produced a prediction of two co-rotating ``clouds'' of plasma, thought to be
responsible for several observed variations, including H$_\alpha$ emission and what appear to be eclipses in the optical photometric light curve.  
Good agreement was found between observations and model computed data, mainly because of the implementation of an offset dipolar magnetic configuration 
to achieve an asymmetry in the size of the two plasma clouds to better fit the observations.  
Regardless, the fit to the photometric light curve by this purely circumstellar model was still unable to reproduce one major feature, a brightening of the star directly after the second cloud passes across the stellar surface.
This discrepancy suggests important physics are missing from the model, and led to a mass effort to obtain new, current data to determine the origin of this feature.  
New and historical photometry demonstrate that the stellar rotation rate is decreasing due to the influence of the magnetic field on angular momentum of the system 
\cite{Townsend:2010aa}.  MOST photometry revealed a precise, stable brightness variation, confirming the asymmetry of the eclipses and the presence of additional brightening unfit by the RRM model
\citep{Townsend:2013aa}.  A set of high resolution spectropolarimetric data of $\sigma$ Ori E was analysed by \citet{Oksala:2012ab} to determine line profile variability and to more precisely calculate 
the variation of the longitudinal magnetic field.  These authors also showed that the de-centered dipole invoked by \citet{Townsend:2005ab} is not compatible with the observed circular polarisation signatures.

In light of these substantial changes in our view of $\sigma$ Ori E, the constraints and parameters used in the RRM model must also be revisited.  Ultimately our goal is to achieve a model such that we properly 
understand the separate contributions to the variability from the star and the magnetically confined material. 
In this work, we present two first steps towards this aim, magnetic Doppler imaging (MDI) and a synthetic light curve analysis.

\section{Magnetic Doppler Imaging\label{MDI}}

We obtained a total of 18 high-resolution (R=65000) broadband (370-1040~nm) spectra of $\sigma$ Ori E 
in November 2007 with the Narval spectropolarimeter on the 2.2m Bernard Lyot 
telescope (TBL) at the Pic du Midi Observatory in France and in February 
2009 with the spectropolarimeter ESPaDOnS on the 3.6-m Canada-France-Hawaii Telescope (CFHT), as part of the 
Magnetism in Massive Stars (MiMeS) Large Program \citep{Wade:2011aa}.

Magnetic Doppler imaging (MDI), previously employed to study Ap/Bp stars, investigates both the surface inhomogeneities of various elements and the magnetic field topology assuming rotational variation of line profiles with time.  
We applied the MDI code {\sc INVERS}10, developed by \citet{Piskunov:2002aa}, to our set of spectra to determine, for $\sigma$ Ori E, the stellar magnetic field 
and surface abundance features of a number of different elements.

\subsection{Magnetic Field}\label{mag}

The Stokes $I$ and Stokes $V$ line profiles of He\,{\sc i} 5867 \AA\,and 6678 \AA\, were used to determine the best fit magnetic topology for $\sigma$ Ori E.
With the longitudinal magnetic field curve computed by \citet{Oksala:2012ab} as an additional constraint in addition to the multipolar 
regularization, the derived magnetic field configuration 
suggests a magnetic field with both a dipole with polar strength $B_{\rm{d}} = 7.434$ kG and orientation angles  $\beta_{\rm{d}} = 47.1^{\circ}$
and $\gamma_{\rm{d}} = 97.5^{\circ}$, and a quadrupole component with strength $B_{\rm{q}} = 3.292$ kG.
The quadrupole axis is offset from the dipole where the positive and negative poles are clearly not separated by 180$^{\circ}$.

\subsection{Surface Abundances}

\begin{figure}[!ht]
\begin{center}
\includegraphics[width=12cm]{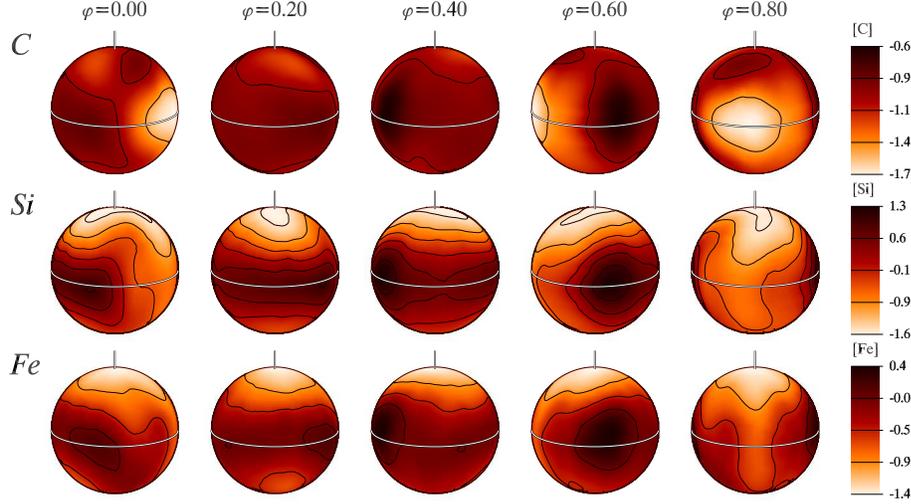} 
\caption{The chemical abundance distribution of $\sigma$ Ori E derived from Stokes $I$ and $V$ profiles of
the C, Fe and Si lines. The star is shown at 5 equidistant rotational phases viewed at the inclination
angle, $i = 75^{\circ}$ and $v \sin i = 140$ km s$^{-1}$. The scale gives abundance as $\epsilon_{Elem}$ corresponding to
$\log(\rm{N}_{Elem} /\rm{N}_{tot})$ for metals. The rotation axis is vertical.  The contour step size is 0.5 dex.}
\label{metalabn}
\end{center}
\end{figure}

\begin{figure}[!ht]
\begin{center}
\includegraphics[width=12cm]{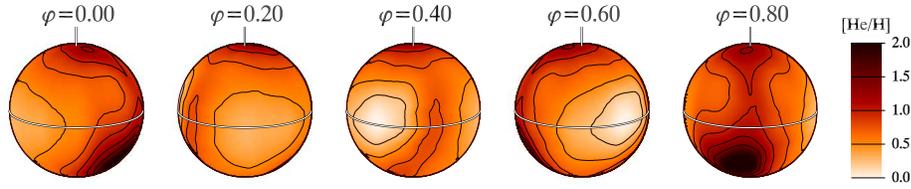}
\caption{The chemical abundance distribution of $\sigma$ Ori E derived from the 4713 \AA\ He\,{\sc i} line. See Fig.\ref{metalabn} for details.
The scale gives abundance as $\log(\rm{He} /\rm{H})$. }
\label{heabn}
\end{center}
\end{figure}

With the derived magnetic field topology set as a fixed parameter, {\sc INVERS}10 was then used to compute the distribution 
of various chemical elements on the surface of the star.    
The set of high resolution time-series spectra of $\sigma$ Ori E contains lines suitable for modeling 
the abundances of Fe, Si, C, and He.  Other chemical elements are present in the spectrum, but their 
line strength was too weak for accurate determination of any surface structure.
He lines required a slightly more rigorous treatment, and were fit with {\sc INVERS}13 \citep[for more details see][]{Kochukhov:2012aa,Kochukhov:2013ab}.

The Stokes $I$ spectral lines were fit to good agreement between the observations and the model fits for all of these lines.  
Figure~\ref{metalabn} displays spherical maps of the abundance distribution for Fe, Si, and C, with He displayed in Figure~\ref{heabn}.   
The maps indicate that the metals all show a similar pattern over the stellar surface.  The minimum abundances are found at 
rotational phase 0.8, in both a spot at the equator and at the top pole.  An equatorial spot at phase 0.5 has the maximum value on the surface.  
The He abundance map indicates a large spot of overabundance at phase 0.8, located in the lower hemisphere.
The minimum abundance is located at phase 0.6, with a normal, solar level of He.  The spot of enhanced He does not appear
to be correlated with the location of the magnetic poles.  Note the anti-correlation between the strength of metals
and He, previously established by \citet{Oksala:2012ab} from EW variations.

\section{Synthesis of Str\"omgren photometric light curves}\label{synlc}

The output from the MDI abundance analysis produces tables of a grid of values over the stellar surface for various latitudes, which
can then be used as input to determine the brightness variation due to the presence of inhomogeneous spots.  \citet{Krticka07} have used this technique
for the case of the He-strong star HD\,37776, and find excellent agreement between synthetic and observed light curves. 
The output of MDI for $\sigma$ Ori E was used as input for a similar treatment, in which grids of TLUSTY model atmospheres \citep{Hubeny:1995aa,Lanz:2007aa} and SYNSPEC synthetic spectra were computed, 
using a fixed $T_{\rm{eff}}$ and $\log g$, but with varying abundance values.

\begin{figure}[!ht]
\begin{center}
\begin{subfigure}{.5\textwidth}
  \centering
  \includegraphics[width=7cm]{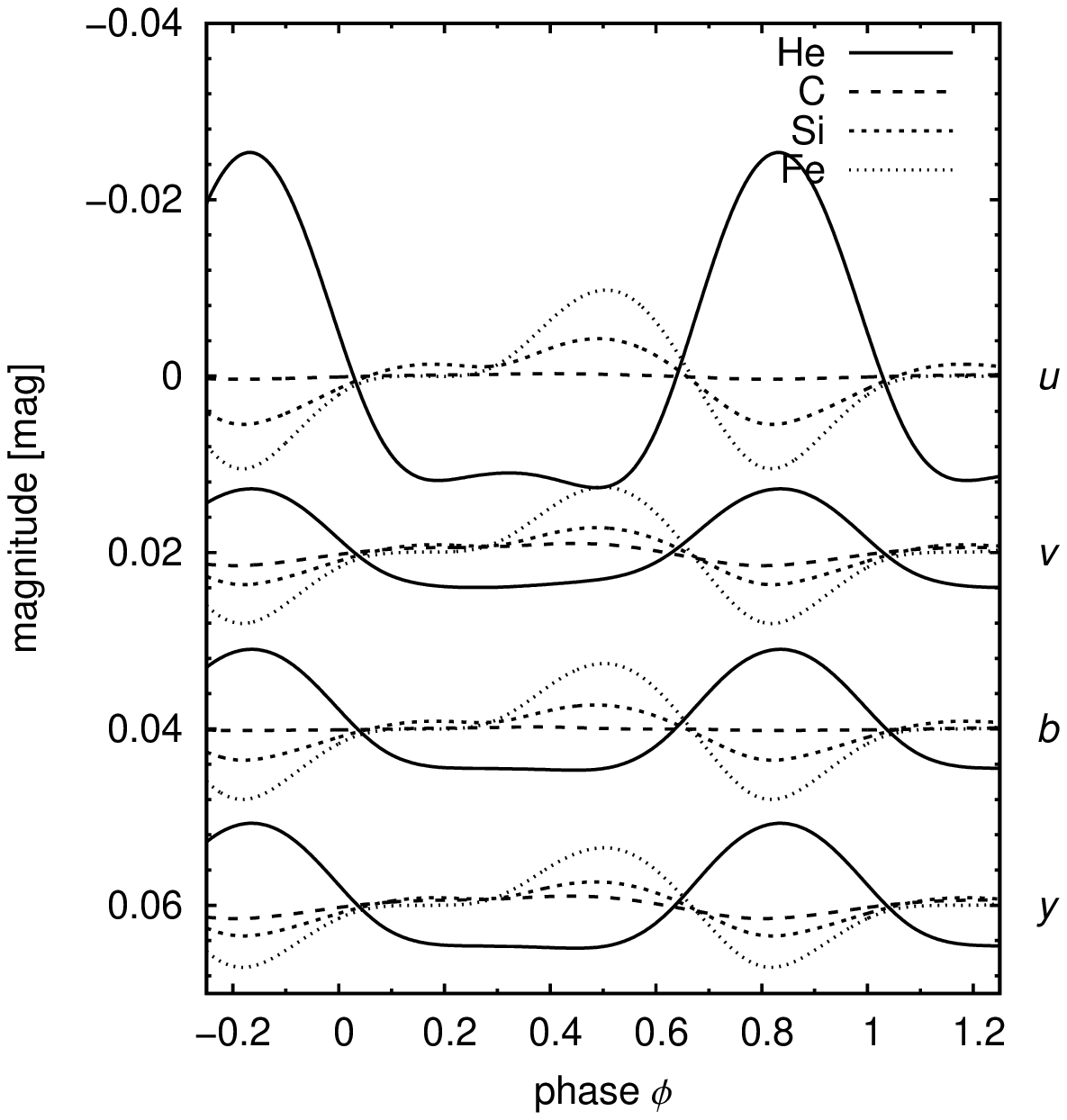}
  \label{fig:sub1}
\end{subfigure}%
\begin{subfigure}{.5\textwidth}
  \centering
  \includegraphics[width=6cm]{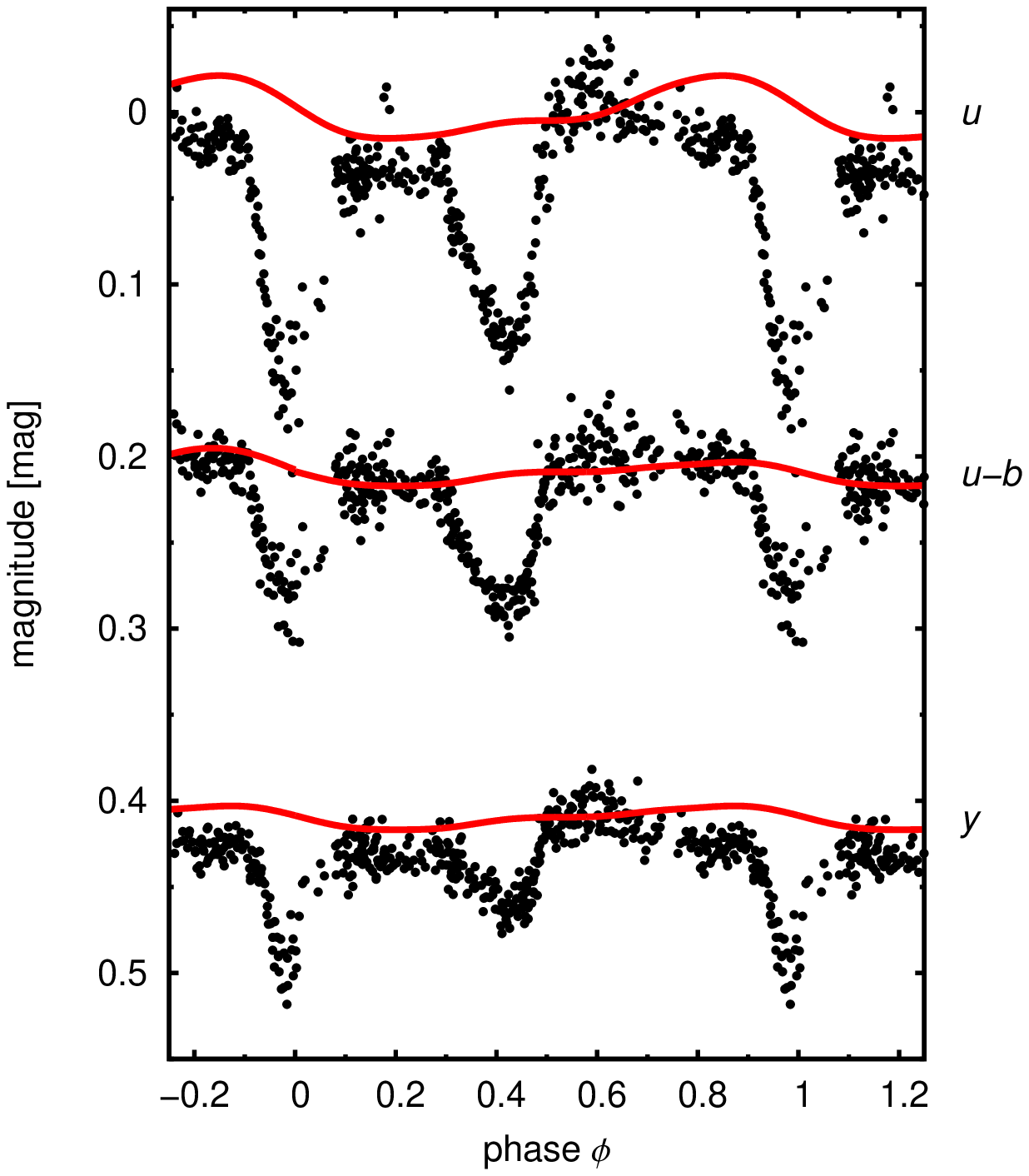}
  \label{fig:sub2}
\end{subfigure}
\caption{\textbf{Left:} Predicted light variations of $\sigma$ Ori E in the Str\"omgren photometric
system calculated using abundance maps of one element only. The abundance of
other elements was fixed. \textbf{Right:} The observed Str\"omgren photometric light curves of $\sigma$ Ori E \citep[black filled dots]{Hesser77}.  Over-plotted is the predicted light variations (red, solid lines) computed taking
into account helium, silicon, carbon, and iron surface abundance distributions.  }
\label{LC}
\end{center}
\end{figure}

The radiative flux in a band $c$, at a distance $D$ from the star, with
radius $R_*$ is \citep{Mihalas:1978aa}
\begin{equation}
\label{vyptok}
f_c=\zav{\frac{R_*}{D}}^2\intvidpo I_c(\theta,\Omega)\cos\theta\,\text{d}\Omega.
\end{equation}
The specific band intensity $I_c(\theta,\Omega)$ is obtained by interpolating between intensities
$I_c(\theta,\varepsilon_\text{He},\varepsilon_\text{Si},\varepsilon_\text{Cr},\varepsilon_\text{Fe})$ 
at each surface point, calculated from the grid of synthetic spectra.
The magnitude difference in a given band is defined as
\begin{equation}
\label{velik}
\Delta m_{c}=-2.5\,\log\,\zav{\frac{{f_c}}{f_c^\mathrm{ref}}},
\end{equation}
where $f_c$ is calculated from Eq.~\ref{vyptok} and
${f_c^\mathrm{ref}}$ is the reference flux obtained under the
condition that the mean magnitude difference over the rotational period is zero.

To obtain predicted light curves, the magnitude differences were computed for each individual rotational phase. For each element, an individual light curve was computed, with the results plotted in the left side of Figure\,\ref{LC}.
Finally, a light curve representative of the photospheric contribution to the brightness variation is produced by combining the effects from all element contributions (right side of Figure\,\ref{LC}).
He dominates this final produce due to its extreme abundance compared to the relatively smaller abundance changes in metals.  As He has its maximum brightness at phase 0.8,
it is clear that the photosphere is not responsible for the excess brightness in the observed photometric light curve seen at phase 0.6.



\section{Conclusions}

We have used new spectropolarimetric observations to study the elemental surface structure and magnetometry of the prototypical Bp star $\sigma$ Ori E.
The analysis and results presented in Section\,\ref{synlc} clearly demonstrate that photospheric inhomogeneities on the stellar surface are not responsible for the
light curve feature not fit by the RRM model.  This conclusion indicates that there are remaining physics missing, likely in the treatment of the circumstellar model, which at present is quite simple.
However, the model for comparison in this case lacks the proper magnetic model determined in Section\,\ref{mag}.  Currently, work is progressing on updating the RRM model to be more consistent with the
observed physical properties of the star.  This work will allow for a more accurate determination of whether the model needs to include additional physical processes, such as scattering.

\bibliographystyle{iau307}
\bibliography{MyBiblio}

\begin{discussion}

\discuss{Khalack}{Could you please comment about possible errors in the simulated light curve caused by taking into account just contribution from horizontal stratification only four elements?}

\discuss{Oksala}{We consider all element contributions from lines that we are able to detect a significant spectral feature with variability. As He dominates the light curve structure, any elements with minimal abundances will not contribute any significant amount.}

\discuss{Tkachenko}{When do you the light curve synthesis, do you assume a global model atmosphere, or you have a grid of models with each model in it being specific to a given surface element?}

\discuss{Oksala}{A grid of models are computed for each combination of varied abundance for all four elements. Each "grid" surface element is correlated with the appropriate model based on the output MDI maps.}

\end{discussion}

\end{document}